\documentclass[twoside,reqno]{HERON}
\usepackage{epsfig,cite,colordvi}
\usepackage{graphicx}
\usepackage{url}
\usepackage{amssymb,amsmath,amscd,epsf}
\usepackage{times}
\usepackage{makeidx}
\begin{document}

\title{Production of strange particles in hadronic interactions}

\runningheads{Production of strange particles in hadronic interactions}
{A. Galoyan,  A. Ribon, V. Uzhinsky}

\begin{start}

\author{A. Galoyan}{1}, \coauthor{A. Ribon}{2}, \coauthor{V. Uzhinsky}{1,2}

\index{Galoyan A.}
\index{Ribon A.}
\index{Uzhinsky V.}

\address{JINR, Dubna, Russia}{1}

\address{CERN, Geneva, Switzerland}{2}

\begin{Abstract}
The NA61/SHINE collaboration has recently published high precision
data on production of $\pi^\pm$ and $K^\pm$ mesons, protons,
antiprotons and $\Lambda$ hyperons  in {\rm pp} interactions at
20, 31, 40, 80 and 158 GeV/c, and in {\rm pC} interactions at 31
GeV/c. The collaboration also presented experimental data on
production of particles -- $\pi^\pm$, $K^\pm$, $p^\pm$, $\rho^0$, $\omega$
and $K^{*0}$ in $\pi^-{\rm C}$ collisions at 158 and 350 GeV/c.
The collaboration has compared these data with various Monte Carlo
model calculations: UrQMD, EPOS, GiBUU, and others.
 All of the models have various problems.
The latest version of the FTF (Fritiof) model of Geant4 solves most of these problems.
In the FTF model, we have improved the fragmentation
of quark-gluon strings with small masses and introduced
dependencies of probabilities of strange mesons and baryon-antibaryon
pair's creation on string masses. Due to these changes, we describe the data
of the NA61/SHINE collaboration on particle production in {\rm pp,
pC}, and $\pi^-${\rm C} interactions.

The improved Geant4 FTF model also well reproduces  experimental
data on inclusive cross sections of $\Lambda, \bar{\Lambda}$ and
$K^{0}$ production in antiproton-proton interactions at various
energies. The modified  FTF model allows one to simulate realistic
processes with two particle productions -- $\bar{p}p \rightarrow
\Lambda \bar{\Lambda}$,~~ $\bar{p}p \rightarrow K^{+} K^{-}$,~~
$\bar{p}p \rightarrow \Lambda \bar{\Sigma}$, and $\bar{p}p
\rightarrow\Sigma \bar{\Sigma}$, which will be studied in the
future by the PANDA experiment at FAIR (GSI, Germany).
\end{Abstract}
\end{start}

\section{Introduction}

Recently, very interesting and useful experimental data were published by the NA61/SHINE
collaboration \cite{NA61}.
In 2017, the NA61/SHINE collaboration presented detailed experimental data on $\pi^{\pm}$, 
$K^{\pm}$, protons and antiprotons production in proton-proton interactions in a wide energy 
range from 20 GeV/c to 158 GeV/c \cite{NA61-PP}.
 In 2016,  experimental data on $\Lambda$ hyperon production
in p-p interactions at 158 GeV/c \cite{NA61PP-Lambda} were published by the collaboration.
At the beginning of 2016, the spectra of $\pi^{\pm}$, $K^{\pm}$, $K^{0}_{s}$, 
$\Lambda$ and protons
in proton-carbon interactions at initial momentum 31 GeV/c \cite{NA61-PC} appeared.
Recently, the NA61/SHINE collaboration published a paper about the production of
various hadrons in pion-carbon interactions at two initial momenta 158 GeV/c and 
350 GeV/c \cite{NA61-piC}.

Authors of the papers performed a comparison of the measured experimental data  with
predictions of various Monte Carlo models -- EPOS 1.99 \cite{EPOS}, 
UrQMD 3.4 \cite{URQMD1} \cite{URQMD2}, GiBUU 1.6 \cite{GiBUU1}, \cite{GiBUU2} 
and Geant4 \cite{Geant4} hadronic models: FTFB and QGSM.
 Main problems were observed in model descriptions of strange particles production.

The aim of our work was to improve the Geant4 FTF model to reproduce 
inclusive spectra of strange particles.
 The Geant4 FTF model is based on the Fritiof model \cite{Fritiof1}, \cite{Fritiof2} of 
 the LUND university.
 We have modified the FTF model using  the experimental data on strange particle 
 and antiproton productions
 in proton-proton interactions at initial momenta larger than 20 GeV/c \cite{NA61-PP}.
 Then, we used the improved FTF model for describing the experimental data on 
 $p{\rm C}$ and ${\pi^-{\rm C}}$
 interactions given in the papers \cite{NA61PP-Lambda}, \cite{NA61-PC}, \cite{NA61-piC}.
 We also used the new FTF model to reproduce  differential cross sections of $\Lambda$-$\Bar \Lambda$ 
 and $K^{+}$-$K^{-}$ productions in antiproton-proton interactions \cite{barnes}, \cite{Jayet}, 
 \cite{Becker}, and kinematical properties of $\Lambda$, $\Bar \Lambda$, $K^{0}_{S}$ produced in 
 antiptoton-proton interactions  in a wide initial energy range -- from 360 MeV/c up to 100 GeV/c 
 \cite{Banerjee}, \cite{Cooper}, \cite{Bertrand},  \cite{Batyunya}, \cite{Ward}.

\section{Inclusive spectra of particles produced in proton-proton interactions at momenta 
of 20, 31, 40, 80 and 158 GeV/c}

It is assumed in the FTF model that hadrons turn into excited states 
in hadron-hadron inelastic interactions.
If only one hadron is excited, the process is called diffraction dissociation. 
The excited states of hadrons are considered as quark-gluon strings.
The Lund fragmentation model is used for  quark-gluon string fragmentation.  
Strange particles are produced by the fragmentation of quark-gluon strings 
containing a strange quark or antiquark coming from the creation of
$s \Bar s$ pairs from the vacuum.   

We have introduced some improvements in the FTF model for a better
description of the experimental data \cite{NA61-PP} on $\pi^{\pm}$, proton, 
antiproton, and $K^{\pm }$ productions in $pp$-interactions at 
$P_{lab}$ $\geq$ 20 GeV/c. 

\begin{figure}[h]
\centering
\includegraphics[scale=0.45]{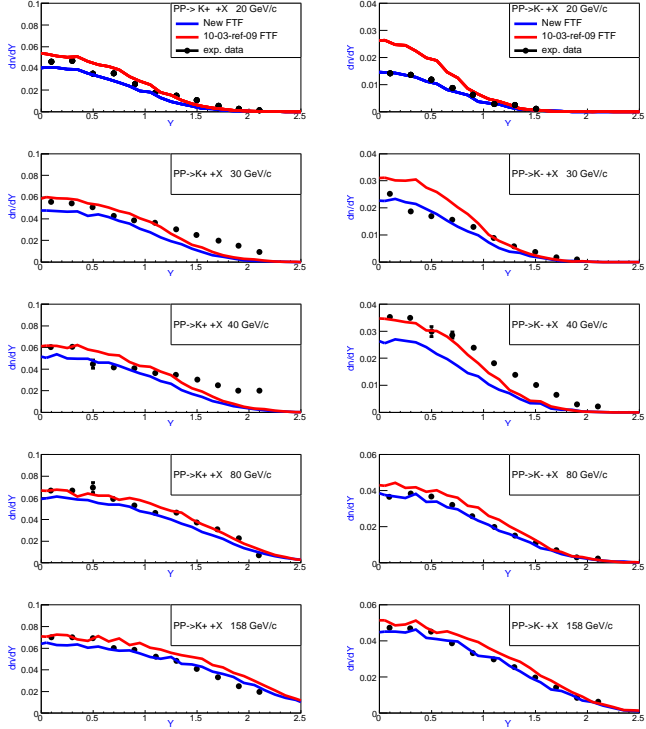}
\caption[]{Rapidity distributions of $K^{+}$ (left side) and $K^{-}$ (right side) mesons 
in pp interactions at various momenta.}
\label{f1}
\end{figure}

In Figure 1, the rapidity distributions of $K^{+}$ and $K^{-}$ mesons 
produced in $pp$-interactions at momenta 20, 30, 40, 80 and 158 GeV/c
 are given. 
The experimental data are presented by points, the red lines are 
calculations with the FTF model in the Geant4 10.03.09 development release. 
As one can see, the calculations
overestimate the K meson production at low energy, 
and describe the data at high energies.

The overestimation at low energies is connected with too large probability of 
strange quark production at string decays. We suppressed the 
probability of string $q \Bar q$ pair production according to the formula:
\begin{equation}
\label{myeq1}
P_{\Bar s s} = 0.108 \cdot (1-(m_{th}/M_{str})^4) ,
\end{equation}
\noindent where 0.108 is an asymptotical value of the probability, $m_{th}$ is a threshold 
mass equals to the sum of
two strange meson masses, $M_{str}$ is a string mass. At this probability, 
the improved FTF model of Geant4
describes well the experimental data at all energies. Only at 40 GeV/c 
the new FTF model cannot describe $K$ meson
spectra. We note that these experimental data do not seem to follow the
common trend of all the rest of the data.

In Figure 2, rapidity distributions of protons and antiprotons
produced in $pp$-interaction at momenta 20, 30, 40, 80 and 158 GeV/c \cite{NA61-PP}
 are shown in a comparison with FTF model calculations. 
Looking at the antiproton spectra calculated by the FTF model of Geant4 10.03.09 release,
it is seen that the calculations  overestimate antiproton production at low energies. 
To suppress this over production,
we introduced a probability of diquark-antidiquark pair creation at a string decay
 given by the formula:
\begin{equation}
\label{myeq2}
P_{q q - \Bar q \Bar q} = 0.15 \cdot (1-(1400 \cdot N_{b}/M_{str})^{2}) ,
\end{equation}

\noindent where 0.15 is the asymptotical value of the probability, $M_{str}$ is the string mass,
and $N_{b}$ is a maximum number of baryons and antibaryons what can be produced. The FTF model 
calculations  with
this improvement are shown in the figure by the blue lines. As seen, the calculations reproduce 
well the antiproton
distributions. It is very important that these changes of the FTF model don't affect the proton 
inclusive spectra.
The FTF model describes the proton rapidity distributions quite well. The same situation 
is observed for $\pi$ meson spectra.

\begin{figure}[h]
\centering
\includegraphics[scale=0.4]{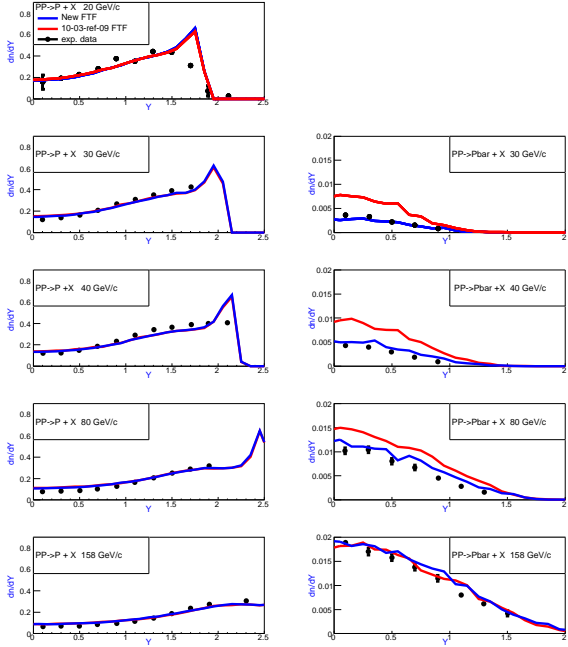}
\caption[]{Rapidity distributions of protons (left side) and antiprotons (right side) 
in pp interactions at various momenta.}
\label{f2}
\end{figure}

The NA61/SHINE collaboration also published various specrtra of $\Lambda$ hyperons, 
especially, rapidity and Feyman X
distributions \cite{NA61PP-Lambda} in proton-proton interactions at $P_{lab}$=158 GeV/c.
The authors of the paper compared these data with various model calculations -- UrQMD, 
Fritiof 7.0 and EPOS 1.99. Only EPOS model described the experimental data. 
Now, the FTF model also describes well these data, as seen in Figure 3.

\begin{figure}[h]
\centering
\includegraphics[scale=0.3]{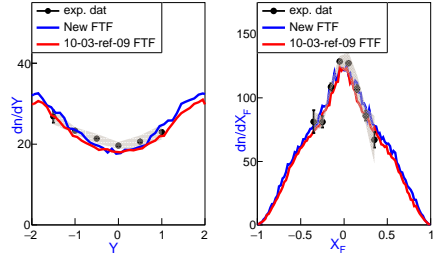}
\caption[]{Rapidity  (left side) and Feyman X (right side) distributions of 
$\Lambda$-hyperons  in pp interactions at 158 GeV/c.}
\label{f3}
\end{figure}
We conclude that the improved FTF model works as well as the EPOS 1.99 model.

\section{Production of strange particles in hadron-nucleus interactions}

The FTF model uses the Glauber approach for estimating the number of participating 
hadrons in hadron-nucleus collisions. In the model, the participating hadrons transform
into quark-gluon strings which fragment into secondary particles, including the 
strange particles.
 
The  NA61/SHINE collaboration presented measurements of spectra of charged pions, 
kaons, and protons as well as $K^{0}_{s}$ and $\Lambda$
hyperons in $p{\rm C}$ interactions at 31 GeV/c in the paper \cite{NA61-PC}. 
In that paper, data are compared with calculations of FTF + Binary Cascade model, 
Venus, EPOS1.99, and  GiBUU models. The conclusion of the paper was that none of
the models provided a satisfactory description of all the spectra. 

\begin{figure}[h]
\centering
\includegraphics[scale=0.4]{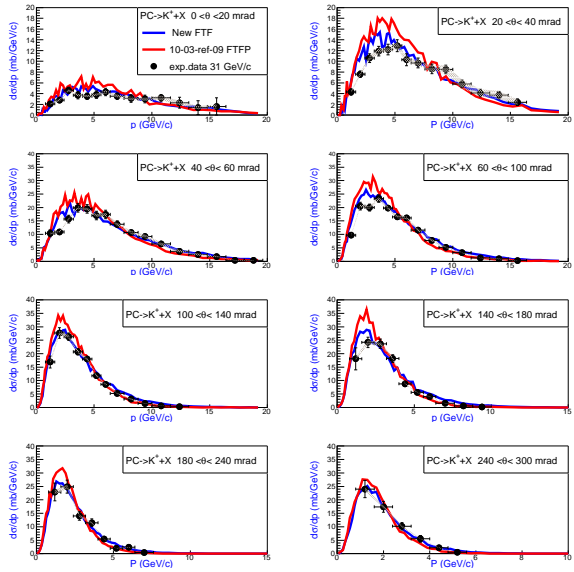}
\caption[]{Momentum distributions of $K^{+}$ mesons in $p{\rm C}$ interactions at 
31 GeV/c at various angles.}
\label{f4}
\end{figure}

In Figure 4, we present the differential cross sections of
$K^{+}$ mesons produced in $p{\rm C}$ interactions as functions of laboratory
 momentum of the mesons. The experimental data  are given at
various emission angles of $K^{+}$  -- 0-20 mrad, 20-40 mrad,
40-60 mrad and so on up to 240-300 mrad.
 Before the improvement, the FTF model calculations (red lines) overestimated  the maxima of the 
distributions by about 50$\%$.
After the improvement, the FTF model reproduces the inclusive spectra of $K^{+}$ mesons quite well.
The same situation takes place for the momentum distributions 
of $K^{-}$ and $K^{0}_{S}$ mesons produced in $p{\rm C}$
interactions at 31 GeV/c. Now, we have reached a good agreement between the  
FTF calculations and the experimental inclusive spectra of mesons and $\Lambda$ hyperons.

Very interesting experimental data on $\pi^{-}{\rm C}$ interactions have been published by 
the NA61/SHINE collaboration in Ref. \cite{NA61-piC}.

\begin{figure}[h]
\centering
\includegraphics[scale=0.4]{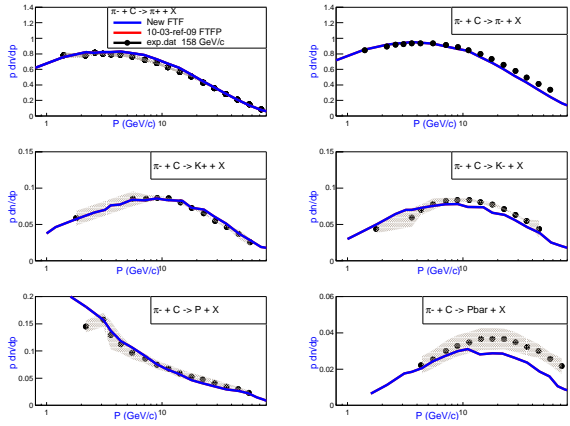}
\caption[]{Momentum distributions of $\pi^{+}$, $\pi^{-}$, $K^{+}$, $K^{-}$, 
$p$, $\Bar p$ in $\pi^{-}$C interactions at 158 GeV/c.}
\label{f5}
\end{figure}

Figure 5 shows the experimental data on $\pi ^{\pm}$  , $K^{\pm}$ , proton and antiproton momentum 
distributions in $\pi^{-}{\rm C}$ interactions at 158 GeV/c. The authors of the paper \cite{NA61-piC} 
showed also calculations of different models: QGSJet, Sybill, DpmJet and EPOS.
The main difference between the models was observed in the calculations of $K^{+}$, $K^{-}$ mesons 
and antiprotons. In Figure 5, we present the FTF model calculations.
In general, the improved FTF model calculations (blue lines) describe better the experimental 
data than other models presented in Ref. \cite{NA61-piC}.
The FTF model results are in agreement with the data on $\pi^{\pm}$, protons, and $K^{\pm}$ mesons. 
At the same time, the FTF calculations are slightly lower than 
the data for antiprotons with momenta larger than 10 GeV/c.

We have a similar situation for the description of the experimental data on pion, kaon, proton and 
antiproton production in $\pi^{-}{\rm C}$ interactions at 350 GeV/c \cite{NA61-piC}.

Summing up, we  conclude that the FTF model with the new probabilities describes quite well
the inclusive spectra of protons, antiprotons, $\pi^{\pm}$, $K^{\pm}$,  $K^{0}$ mesons 
and $\Lambda$-hyperons produced in $pp$, $p{\rm C}$, and $\pi^{-}{\rm C}$ interactions.

\section{Production of strange particles in antiproton-proton interactions}

A detailed experimental study of antiproton-proton and antiproton-nucleus interactions 
is foreseen at the future FAIR facilities (GSI, Darmstadt, Germany) by the PANDA 
Collaboration \cite{Panda}.
Simulations of antiproton  interactions with protons and nuclei
are needed for cosmic space experiments like PAMELA, BESS, AMS, CAPRISE,
which are going to search for anti-nuclei in cosmic rays.
For the simulation of antiproton-proton interactions, the Geant4 FTF
model uses the main assumptions of the Quark-Gluon-String Model \cite{DPM}. 
The FTF model assumes production and fragmentation of quark-anti-quark and
diquark-anti-diquark strings in the mentioned interactions. Main ingredients of
the FTF model are cross sections of string creation processes
\cite{Baldin} and the use of the LUND
string fragmentation algorithm. They allow to achieve a satisfactory description of
a large set of experimental data, in particular, strange particle production,
such as $\Lambda$ hyperons and $K$ mesons \cite{bolg16}.
To develop the hypernuclei program of the Panda experiment, it is needed
to estimate multiplicity of hyperons and correctly reproduce their 
kinematical characteristics.
Thus, we calculate with the FTF model the average multiplicities of $\Lambda$ 
hyperons and $K^0_S$ mesons in anti-proton-proton interactions presented in Figure 6.

\begin{figure}[h]
\centering
\includegraphics[scale=0.4]{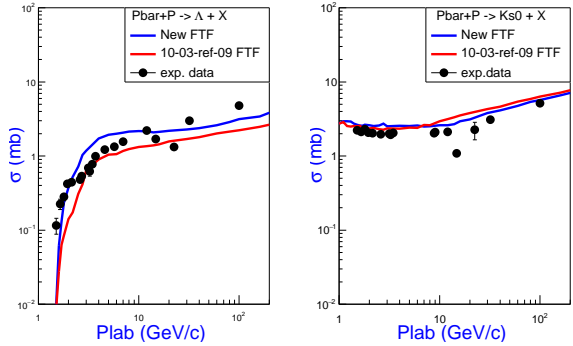}
\caption[]{Inclusive cross sections of $\Lambda$-hyperon and  $K^{0}_{s}$-meson production in
antiproton-proton interactions as function of projectile momentum.}
\label{f6}
\end{figure}

The Figure shows experimental data \cite{Banerjee} on inclusive cross section of 
$\Lambda$ hyperon and $K^0_S$ meson
in antiproton-proton interactions as a function of projectile momentum.
As seen, the FTF model describes well $K^0_S$ meson cross sections at low energies,
and overestimates, a little bit, the yield of $K^{0}_{S}$ mesons at momenta more than 10 GeV/c.
Calculations by the old version of FTF underestimated the production cross section 
of $\Lambda$ hyperons at low momenta, less than 2 GeV/c, and were in agreement with 
the experimental cross section at higher energies up to 30 GeV/c.
The developed FTF model calculations are in agreement with $\Lambda$ production 
cross section at low energies up to 3 GeV.

Distributions of the strange particles on kinematical variables
 are very important for experiments. Distributions of $\Lambda$
hyperons and $K^0_S$ mesons on rapidities, Feyman X, and
squared transverse mometum in $\Bar pp$ interactions at momenta
of 12, 24 and 100 GeV/c calculated with the FTF model are
presented in Figure 7. 

\begin{figure}[h]
\centering
\includegraphics[scale=0.5]{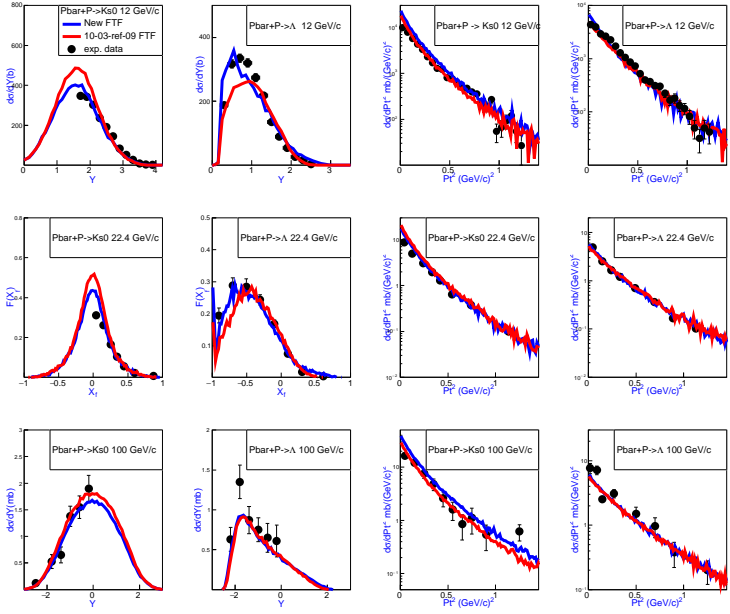}
\caption[]{Rapidity, Feyman X, $Pt^{2}$ distributions of $K^{0}_{s}$-mesons and 
$\Lambda$-hyperon in $\Bar p p$ interactions at 12, 22.4, 100 GeV/c.}
\label{f7}
\end{figure}

As seen, there is a good agreement between
the improved FTF model calculations (blue lines) and the
experimental data on the inclusive spectra at  12 GeV/c
\cite{Bertrand}  and 22.4 GeV/c \cite{Batyunya}.

When we consider antiproton-proton interactions at very high momenta, 100 GeV/c, annihilation 
processes are absent.
Here, only diffraction processes of FTF are working. At this energy, FTF gives a
 quite good description of $K^0_S$ and $\Lambda$ rapidity and $P_{T}^2$ spectra.

Let us consider now binary reactions with strange particle production in  antiproton-proton annihilations.

There are a lot of experimental  data on differential cross sections of $\Lambda$-$\Bar \Lambda$ 
production in antiproton-proton interactions. We have gathered most of them, 
performed corresponding simulations by the FTF model and compared with
the experimental data. For example, in Figure 8 we present differential cross sections 
of $\Lambda$-$\Bar \Lambda$ at various initial momenta: 1.56, 1.76, 1.88 and 2.06 GeV/c.  

\begin{figure}[h]
\centering
\includegraphics[scale=0.5]{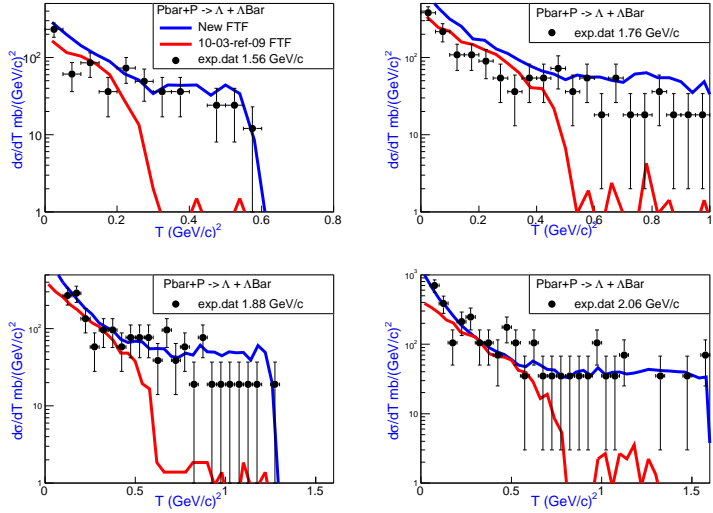}
\caption[]{Differential cross section of $\Lambda$-hyperon production as 
function of transferred momentum in $\Bar p p$ interactions at various energies.}
\label{f8}
\end{figure}

To describe the data \cite{Jayet}, we introduced in the FTF model the spatial rotation of quark-gluon 
strings.
The FTF model without rotating strings does not describe the experimental data \cite{Jayet}. 
The FTF model results with the rotating strings are in agreement with the experimental distributions.

\section{Conclusion}
A new formula for the probability of strange $q \Bar q$  pair production at string  decay is 
proposed and implemented in the FTF model of Geant4.
The new probability improves considerably the description of the strange particle production in 
the FTF model.

A new formula for the probability of diquark-antidiquark pair production at string decay is  
proposed and implemented in the FTF model.
The new pro\-ba\-bi\-li\-ty helps to improve the description of antiproton and $\Lambda$ hyperon 
production in the FTF model.

Good agreement between the improved FTF model calculations and 
the NA61/ \\ SHINE experimental data 
is reached on  $K^{\pm}$, $K^{0}_{S}$ , and $\Lambda$ production in proton-proton, proton-Carbon and  
$\pi^{-}$-meson-Carbon collisions in a wide energy range.

Kinematical properties of $\Lambda$ hyperons and $K^{0}_{S}$ mesons produced in $\Bar p  p$ reactions 
are calculated in the improved FTF model with the rotating strings, and compared with the 
experimental data at various initial momenta. 
Good description of the experimental data is obtained in the FTF model with the new probabilities.

The differential cross  sections of the processes  $\Bar p p \to
\Lambda \Bar \Lambda$ are calculated in various versions of the
FTF model. Reasonable description of the experimental data is
reached in the FTF model with the rotating strings and the new pro\-ba\-bi\-li\-ti\-es.

\section{Acknowledgements}
A. Galoyan is thankful to heterogeneous computing team of LIT
JINR (HybriLIT) for support of calculations  and to Professor Cheuk-Yin Wong for 
helpful discussions.


\begin{thebibliography}{99}
\bibitem{NA61} URL:\url{https://home.cern/about/experiments/na61shine}.

\bibitem{NA61-PP} NA61/SHINE Collaboration, \textit{Eur.Phys.J.} \textbf{C77} (2017) no.10, 671.

\bibitem{NA61PP-Lambda} NA61/SHINE Collaboration \textit{Eur. Phys. J.} \textbf{C76} (2016) no.4, 198.

\bibitem{NA61-PC} NA61/SHINE Collaboration  \textit{Eur. Phys. J.} \textbf{C76} (2016) no.2, 84

\bibitem{NA61-piC} NA61/SHINE Collaboration (Raul R. Prado (Sao Paulo U., Sao Carlos)
Jul 25, 2017. \textit{Conference: C17-07-12 Proceedings}. \textit{arXiv:1707.07902} [hep-ex]

\bibitem{EPOS}K. Werner, \textit{Nucl. Phys. Proc. Suppl.}  175-176 (2008) 81-87.
\bibitem{URQMD1}S. Bass et al., \textit{Prog.Part.Nucl.Phys.} \textbf{41} (1998) 255, 
\textit{arXiv:9803035} [nucl-th].
\bibitem{URQMD2}M. Bleicher et al., \textit{J.Phys.} \textbf{G25} (1999) 1859, 
\textit{arXiv:9909407} [hep-ph].
\bibitem{GiBUU1}  O. Buss, T. Gaitanos, K. Gallmeister, H. van
    Hees, M. Kaskulov, et al., \textit{Phys. Rept.}
    \textbf{512} (2012) 1.
\bibitem{GiBUU2} K. Gallmeister, U. Mosel, \textit{Nucl. Phys.}
    \textbf{A826}  (2009) 151.
\bibitem{Geant4}J.Allison et al., \textit{Nucl.Instrum.Meth.}
    \textbf{A835} (2016) 186.
\bibitem{Fritiof1} B. Andersson et al., \textit{Nucl. Phys.}
    \textbf{B281} (1987) 289.
\bibitem{Fritiof2} B. Nilsson-Almquist and E. Stenlund,
    \textit{Comp. Phys. Commun.} \textbf{43} (1987) 387.
\bibitem{barnes} P.D. Barnes et al., \textit{Nuclear Physics}
    \textbf{A558} (1993) 277.
\bibitem{Jayet}  B. Jayet et al., \textit{Nuov. Cim.}
    \textbf{A45}  (1978) 371.
\bibitem{Becker}  H. Becker et al., \textit{Nucl. Phys.}
    \textbf{B141} (1978) 48.
\bibitem{Banerjee} S. Banerjee et al., \textit{TIFR-BC-78-8}
\bibitem{Cooper}  A.M. Cooper et al., \textit{Nucl.Phys.}
    \textbf{B136} (1978) 365.
\bibitem{Bertrand} D. Bertrand et al., \textit{Nucl. Phys.}
    \textbf{B128} (1977) 365.
\bibitem{Batyunya}  B.V. Batyunya et al., \textit{Z. Phys.}
    \textbf{C25} (1984) 213.
\bibitem{Ward} D.R. Ward et al., \textit{Phys. Lett.}
    \textbf{62B} (1976) 237.
\bibitem{Panda} PANDA Collaboration (M.F.M. Lutz et al.) \textit{arXiv:0903.3905} [hep-ex] (2009). 
\bibitem{DPM} A.~Capella, U.~Sukhatme, C-I~Tan,
    J.~Tran~Thanh~Van, \textit{Phys. Rept.} \textbf{236} (1994)
    225; A.B.~Kaidalov and K.A.~Ter-Martirosian, \textit{Phys.
    Lett. B} \textbf{117} (1982) 247.
\bibitem{Baldin}A. Galoyan, A. Ribon and V. Uzhinsky \textit{PoS
    BaldinISHEPPXXII} (2015) 049.
\bibitem{bolg16} A. Galoyan, A. Ribon, V. Uzhinsky \textit{arXiv:1610.08341} [nucl-th] (2016). 

\end{thebibliography}
\end{document}